\documentstyle[12pt,epsf]{article}
\def\m@thcombine#1#2{%
  \setbox0=\hbox{$#1$}
  \setbox1=\hbox{$#2$} 
  \ifdim\wd0>\wd1
    \setbox0=\hbox to\wd1{\hss\box0\hss}
  \else
    \setbox1=\hbox to\wd0{\hss\box1\hss}
  \fi
  \mathop{\vcenter{
    \offinterlineskip\box0\box1}}}
\def\lesim{\m@thcombine<\sim}
\def\gesim{\m@thcombine>\sim}
\def\lessgtr{\m@thcombine<>}
\def\gtrless{\m@thcombine><}
\makeatletter
\def\m@thcombine#1#2{%
  \setbox0=\hbox{$#1$}
  \setbox1=\hbox{$#2$}
  \ifdim\wd0>\wd1
    \setbox0=\hbox to\wd1{\hss\box0\hss}
  \else
    \setbox1=\hbox to\wd0{\hss\box1\hss}
  \fi
  \mathop{\vcenter{
    \offinterlineskip\box0\box1}}}
\def\lesim{\m@thcombine<\sim}
\def\gesim{\m@thcombine>\sim}
\def\lessgtr{\m@thcombine<>}
\def\gtrless{\m@thcombine><}
\def\bold#1{\mbox{\boldmath $#1$}}
\makeatother
\newcommand{\sslash}[1]{\, {\not {\! #1}}}
\begin{document}
\begin{center}
{\bf Longitudinal and Transverse Spin Responses \\
in Relativistic Many Body Theory}
\end{center}
\begin{center}
K. Yoshida and H. Toki 
\end{center}
\vspace{5mm}
\begin{center}
{\it
Research Center for Nuclear Physics (RCNP), Osaka University,
Ibraki, Osaka 567-0047, Japan}
\end{center}

\begin{abstract}
We study theoretically the spin response functions
using the relativistic many body theory.  The spin response functions
in the relativistic theory are reduced largely from the ones of
the non-relativistic theory.  This happens particularly to
the longitudinal spin responses.  This fact is able to remove
the difficulty in reproducing the ratio of the longitudinal and
transverse response functions seen experimentally.  We use the
local density approximation with the eikonal prescription of
the nuclear absorption on the incoming and outgoing nucleons for
the calculations of the response functions of finite nuclei.
We compare the calculated results with the recent experimental results
with $(\vec{p},\vec{n})$ reactions on C and Ca.
\end{abstract}

{\it keywords} : spin response function, relativistic many-body theory,
quasielastic reaction
\section{Introduction}
The spin response functions are very interesting, since they convey
directly the informations on the spin correlations among nucleons.
The longitudinal 
spin responses are related with the pionic correlations, which
have a strong connection with pion condensation and its precritical
phenomenon~\cite{Migdal,Toki}.  The transverse responses are related
with the rho mesonic correlations.  Due to the mass difference between
the pion and the rho meson, the meson exchange force is attractive
in the pion channel while it is repulsive in the rho meson
channel at medium momenta, $q\sim 2$ fm$^{-1}$.  
Hence the comparison of the two spin responses should
reflect this feature directly.  In fact, the ratio between the
spin responses in these channels was predicted to be largely 
different from unity~\cite{Ericson}.

Experiments on the spin responses were performed at LAMPF and RCNP
at intermediate energies following the above ideas in
mind~\cite{LAMPF,RCNP}.  The 
experimental results were found, however, extremely puzzling.
The ratios of the longitudinal and the transverse spin responses
for various nuclei were found close to or less than one, which is
largely different 
from the theoretical expectations~\cite{LAMPF,Wakasa}. The surface
effects were considered by Esbensen {\it et al}., but the ratio could
be brought down only to a factor two~\cite{Esbensen}.  Very involved
calculations were performed 
by Ichimura {\it et al}. by considering the finite nuclear geometry
and the 
realistic distortion effects~\cite{Ichimura}.  The results were
essentially unchanged from that of Esbensen {\it et al}..

These results indicate that something essential is missing in the
spin correlations among nucleons in nuclei.  In this respect, there
was a very interesting observation made by several
authors~\cite{Savushkin,Horowitz}. 
With the use of the pseudovector coupling of the pion with the nucleon
instead of the pseudoscalar coupling in the relativistic framework,
the pion response becomes much weaker than that of the non-relativistic
framework.  The pion response function is reduced by a factor
$(M^*/M)^2$ and the effective mass, $M^*$ reduces largely in nuclear
matter as the density increases in the relativistic many body theory
~\cite{Walecka,Serot,Brockmann,Hirata}.  Hence, even at the normal
matter density without 
the Landau-Migdal short range correlations, i.e., $g'=0$, the pion
condensation 
does not occur~\cite{Savushkin}.  This is a very interesting observation, since
this fact largely affects in the longitudinal spin response, which is
the pion response, and could be the source to remove the discrepancy.

Horowitz and his collaborators are the ones who apply this idea to
the spin response functions and the various spin
observables~\cite{Horowitz}. 
They found that in fact the ratio could be brought down to close to
one in accordance with the experimental observations.  These calculations
were performed in nuclear matter assuming some typical density
representing 
the nucleus of interest.  In this paper, we would like to take
into account the change of the densities by considering the
hadron distortion effects and compare with the experimental data.
In doing so, we would like to compare our results not only with
the ratios of the spin responses, but also with the response functions
themselves.  Particularly, we would like to compare with the
recent intensive experimental studies of the $(\vec{p},\vec{n})$
reactions at 
400MeV by H. Sakai and his collaborators~\cite{Wakasa}.

This paper consists of the following contents.  We write the definition
of the response functions and the expressions of the spin response
functions in the relativistic framework in Sect.2.  We provide the
numerical results in Sect.3, where we demonstrate the effects of
the relativity on the spin response functions both for the
longitudinal and 
the transverse channels at different densities.  We choose here
the matter properties calculated by the relativistic mean field theory
(RMF) with the non-linear terms with the parameter set
TM1~\cite{Sugahara}. 
Sect.4 is devoted to the summary of this work.

\section{Spin response functions}
Spin response functions have been 
studied by the $\pi\!+\!\rho\!+\!g'$ model in the non-relativistic
framework, in which spin longitudinal and spin transverse responses
are induced by the residual p-h interactions described in terms of
$\pi$ and $\rho$ exchange\cite{Oset}.
The $\pi$ exchange involves the coupling vertex $\bold{\sigma}\cdot
\bold{q}$, hence 
gives rise to the longitudinal response with respect to momentum
transfer $\bold{q}$, and the $\rho$ exchange involves
$\bold{\sigma}\!\times 
\!\bold{q}$ which induces the spin transverse response.
The relativistic version of the spin responses is described in the 
Walecka model~\cite{Walecka}
with  the 
spin dependent interaction Lagrangian~\cite{Brockmann} 
\begin{equation}
\label{eq:int}
{\cal L}_{\rm int}=\frac{f_\pi}{m_\pi}\bar{\psi}_N \gamma_\mu\gamma_5
\bold{\tau}\psi_N \partial^\mu \bold{\pi} - g_\rho \bar{\psi}_N
(\gamma_\mu+\frac{\kappa}{2M}\sigma_{\mu\nu}\partial^\nu)\bold{\rho}_\mu
\bold{\tau}\psi_N .
\end{equation}
$M$ is the nucleon mass and $\kappa\!=\!6.6$ is the tensor-to-vector
ratio in 
the $\rho NN$ coupling~\cite{Hohler}. The $\pi NN$ coupling is assumed 
to be of the pseudovector representation which incorporates the
correct low-energy pion behavior~\cite{Matsui}.

The first analysis of the spin responses in the relativistic
 model was done by
Horowitz {\it et al}., who calculated spin 
observables~\cite{Horowitz}. Our goal is to 
reveal the relativistic effects on the spin response 
functions and compare with the experimental cross
sections observed in $(\vec{p},\vec{n})$ reactions on
finite nuclei, with nuclear distortion effect taken into account.
The spin response functions are obtained by taking the imaginary part
of the polarization functions induced by the interaction Lagrangian
(\ref{eq:int}) as functions of the 4-momentum transfer
$q=(\omega,\vec{q})$. 
For nuclear matter the pseudovector and $\rho$ meson induced
polarization functions in the lowest order are given as  
\begin{eqnarray}
i\Pi_{\rm PV}(q)&=&{\rm Tr}\int
\frac{d^4k}{(2\pi)^4}\gamma_5\!\sslash{q} 
\bold{\tau}_1 
G(k+q)\gamma_5\!\sslash{q} \bold{\tau}_2 G(k)\\\label{eq:polt}
i\Pi_{\mu\nu}(q)&=&{\rm Tr}\int
\frac{d^4k}{(2\pi)^4}\Gamma_\mu 
\bold{\tau}_1 
G(k+q)\bar{\Gamma}_\nu \bold{\tau}_2 G(k)
\end{eqnarray}
where 
$\Gamma_\mu\!=\!\gamma_\mu\!-\!
i\frac{\kappa}{2M}\sigma_{\mu\nu}q^\nu$($\bar{\Gamma}_\mu\!=\!\gamma_\mu
\!+\! i\frac{\kappa}{2M}\sigma_{\mu\nu}q^\nu$). 
$G(k)$ is the nucleon Green's function,
\begin{eqnarray}\nonumber
G(k)&=&G_F(k)+G_D(k)\\\nonumber
G_F(k)=(\!\sslash{\bar{k}}+M^*)g_F(k),&&G_D(k)=
(\!\sslash{\bar{k}}+M^*)g_D(k)\\ \nonumber
g_F(k)=\frac{1}{\bar{k}^2-M^{*2}+i\eta},&&
g_D=\frac{i\pi}{\epsilon_k}\delta(\bar{k}_0-\epsilon_k)
\theta(\epsilon_F-\bar{k}_0)\\\nonumber  
\bar{k}&=&(k_0-\Sigma_V,\bold{k})\\\nonumber
\epsilon_k&=&\sqrt{\bold{k}^2+M^{*2}}.
\end{eqnarray}
$\epsilon_F$ denotes the Fermi energy, $\epsilon_F=\sqrt{k_F^2+M^{*2}}$,
with $k_F$ the Fermi momentum. $M^*=\!M+\Sigma_S$ is the nucleon
effective  mass given by the relativistic mean field theory with
$\Sigma_S,\Sigma_V$ being the scalar and vector potentials.
The transverse polarization
$\Pi_T$ is obtained by projecting
$\Pi_{\mu\nu}$ 
onto the plane perpendicular to $\bold{q}$~\cite{Hatsuda}. Thus 
\begin{eqnarray}\nonumber
\Pi_L&=&\Pi_{\rm PV}\\
\Pi_T&=&(\frac{q_0^2}{\bold{q}^2}\Pi_{00}+\Pi^i_{\,\,i})/2.
\end{eqnarray}
In the calculation of the polarization functions,
we neglect the vacuum polarization of the form $\int
\frac{d^4k}{(2\pi)^4}  
\Gamma G_F\Gamma G_F$
which represents the coupling of the meson to $N\bar{N}$.

The vector current
conservation relation, $q_\mu\bar{\psi}\gamma^\mu\psi=0$, allows us to 
reduce
$\Pi_{\rm PV}(q)$ 
to ~\cite{Dawson}
\begin{equation}
\label{Eq:longitudinal}
i\Pi_{\rm PV}=-16M^{*2}q^2\int \frac{d^4k}{(2\pi)^4}g(k)g(k+q)
\end{equation}
where $g(k)=g_F(k)+g_D(k)$. 
$\Pi_L$ is written with $I_0(q)$ in
~\cite{Kurasawa} as
\begin{equation}
\Pi_L=4M^{*2}q^2I_0.
\end{equation}
The explicit form of $Im \Pi_L$ is given as 
\begin{equation}
\label{Eq:imaginary}
Im \Pi_L=\frac{M^{*2}q^2}{\pi\mid\! \bold{q}\!\mid}
\left \{
 \begin{array}{cc}
\omega &(k_+<k_F)\\\epsilon_F-\epsilon_{k_-} &(\mid\!\! k_- \!\!\mid
<k_F<k_+)\\0 &(k_F<\mid\!\! k_- \!\!\mid)
\end{array} \right .
\end{equation}
and the real part is given as 
\begin{eqnarray}\nonumber
Re \Pi_L&=&\frac{M^{*2}q^2}{\pi\mid\! \bold{q}\!\mid}\left \{
\epsilon_F\ln{\left |
\frac{(k_F-k_+)(k_F-k_-)}{(k_F+k_+)(k_F+k_-)}\right |}
-2\omega\ln{\frac{k_F+\epsilon_F}{M^*}}\right .\\\nonumber
&-&\left . \left (
\epsilon_{k_+} +\ln{\left |\frac{(M^{*2}+k_+
k_F-\epsilon_{k_+} \epsilon_F)(k_F+k_+)}{(M^{*2}-k_+
k_F-\epsilon_{k_+} \epsilon_F)(k_F-k_+)} \right |}
+ (k_+ \rightarrow k_-) \right )
\right \}\\
\end{eqnarray}
with 
\[k_\pm=\frac{\mid\!\bold{q}\!\mid}{2}
\pm\frac{\omega}{2}\sqrt{1-\frac{4M^{*2}}{q^2}}.\]

On the other hand, $\Pi_T$ 
is divided into three components, each of which originates from the
vector-vector, vector-tensor and tensor-tensor contributions, 
\begin{equation}
\Pi_T=\Pi_v+\Pi_{v,t}+\Pi_t,
\end{equation}
\begin{eqnarray}
\Pi_v&=&\frac{1}{2}\left[(4M^{*2}+q^2)I_0+\frac{q^2}{\bold{q}^2}I_2\right]+ 
\frac{\omega^2+\bold{q}^2}{2\bold{q}^2}\frac{\rho_s}{M^*}\\
\Pi_{v,t}&=&4\frac{\kappa}{2M}M^* q^2I_0 \\
\Pi_t&=&\left
( \frac{\kappa}{2M}\right
)^2q^2\left [(4M^{*2}+q^2)I_0-\frac{q^2}{\bold{q}^2}
I_2-\frac{q^2}{\bold{q}^2} \frac{\rho_s}{M^*}\right ].
\end{eqnarray}
$\rho_s$ is the scalar density 
and 
$I_0$ and $I_2$ are given 
in ~\cite{Kurasawa}, where the response
functions in $(e,e')$
inelastic scattering were calculated. 

In order to define the response functions in the relativistic model, it is
necessary to consider the non-relativistic limit of the vertex operators
since all the discussions have been made in the non-relativistic
framework.
While $\gamma_5\gamma_{\mu}q^\mu \rightarrow \bold{\sigma}\cdot
\bold{q}$,
\begin{equation}
\gamma_\mu
-i\frac{\kappa}{2M}\sigma_{\mu\nu}q^\nu\rightarrow \frac{\kappa
+1}{2M}\bold{\sigma}\!\times \!\bold{q},
\end{equation}
in the non-relativistic reduction.
Thus, we define the relativistic response functions $R_L$, $R_T$ by
\begin{eqnarray}
\label{eq:definition}
q^2 R_L&=&\frac{1}{\pi\rho}Im \Pi_L\\
\left ( \frac{\kappa+1}{2M}\right )^2q^2R_T&=&\frac{1}{\pi\rho}Im \Pi_T,
\end{eqnarray}
where $\rho$ is the nuclear density.
They correspond to 
the response functions in non-relativistic model $R_{\rm NR}$ defined by
\begin{equation}
\bold{q}^2R_{\rm NR}=-\frac{1}{\pi\rho}Im \Pi_{\rm NR}
\end{equation}
where $\Pi_{\rm NR}$ are the non-relativistic polarization functions 
for the vertex operators,
$\bold{\sigma}\cdot \bold{q}$ or $\bold{\sigma}\!\times \!\bold{q}$.
We remind you that in the non-relativistic model
the longitudinal response and the transverse
response are identical for the symmetric nuclear matter and they are
expressed by the  
Lindhard function given in ~\cite{Walecka-text}. 

Here we give an overview of the relativistic responses in comparison
with the non-relativistic ones.
First we discuss the longitudinal response functions.
In the high density case where $k_F$ is large, the bulk part of
the response function is given by the linear term in $\omega$,
the first row in Eq.(\ref{Eq:imaginary}). A typical case is shown in
Fig.1.  
We find that in the linearly $\omega$-depending region 
we gain an exact relation,
$R_L=\left ( \frac{M^*}{M}\right )^2\! R_{\rm NR}$.
In the low density case where $k_F$ is small, most part of the
response function is represented by 
the second row in Eq.(\ref{Eq:imaginary}).
This expression differs from the non-relativistic response only in the
relativistic kinematics and the effective mass.
We find that at the peak position
$\omega=\sqrt{\bold{q}^2+M^{*2}}-M^*$ obtained by setting $k_-\!=0$ in
Eq.(\ref{Eq:imaginary}), $-Im\Pi_L=-\frac{M^{*2}q^2}{\pi\mid
\bold{q}\mid}(\epsilon_F-M^*)\approx
-\frac{M^{*}k_F^2q^2}{2\pi\mid 
\bold{q}\mid}$ shows approximately linear dependence on the effective
mass.
In the non-relativistic limit with $M^*=M$ this can be reduced to the 
non-relativistic case which provides the 
peak position 
$\omega=\frac{\bold{q}^2}{2M}$ and the peak height  
$-Im\Pi_{\rm NR}=\frac{Mk_F^2\bold{q}^2}{2\pi \mid \bold{q}\mid}$. 
This implies 
that the relativistic responses are reduced by a factor $\left
( \frac{M^*}{M}\right )^2$ at higher density and by
$\frac{M^*}{M}$ at lower density compared with the
non-relativistic one at the 
same density. 

Second we discuss the transverse response.
$\Pi_T$ is rather complicated 
to make a clear and brief analysis over all $\omega$ region.
Hence, we analyze here the high density case where at lower $\omega$
\begin{eqnarray}
Im I_0&=&\frac{\omega}{4\pi \mid \bold{q}\mid}\\
Im I_2&=&\frac{1}{12\pi\mid
\bold{q}\mid}(12(M^{*2}+k_F^2)\omega+\omega^3)\\
&\approx& 4(M^{*2}+k_F^2) Im I_0.
\end{eqnarray}
The leading order of $Im I_0$ and $Im I_2$ in $M^*$ yields
\begin{eqnarray}
Im \Pi_T&\approx& \left
[ -2k_F^2+\frac{1}{2}q^2+2\kappa\frac{M^*}{M}q^2+\kappa^2\left
(\frac{M^*}{M}\right )^2q^2 \right]Im I_0\\
&\sim& \left (1+\kappa\frac{M^*}{M}\right )^2q^2 Im I_0.
\end{eqnarray}
We can find by considering the definition Eq.(\ref{eq:definition})
that setting $M^*=M$ makes $R_T$ approximately agree with $R_L$ and
therefore 
with $R_{NR}$.
Pay attention to the fact that $R_T$ depends on $M^*$ in more
moderate way than $R_L$. 
This arises from the vector-tensor mixed coupling of $\rho$ to
nucleons;
the trace appearing in the calculation of $\Pi_{v,t}$ 
in Eq.(\ref{eq:polt}) does not give rise to $M^{*2}$
term, {\it i.e.}, 
$Tr[\gamma_\mu (\sslash{k}+\sslash{q}+M^*)\sigma_{\nu\lambda}
(\sslash{k}+ M^*)]q^\lambda=4M^*(g_{\mu\nu}q^2-q_\mu q_\nu)$ due to
the vanishing traces of product of odd numbers of Dirac $\gamma$
matrices, while 
$M^{*2}$ term appears in the other. 
This leads to $R_L/R_T$ smaller than unity with $M^*/M<1$ as
illustrated in Fig.1 where $R_L$ is found smaller than $R_T$ in the
whole $\omega$ region. We see clearly the difference between $R_L$ and 
$R_T$. Here, we have used the effective nucleon mass $M^*$ by using
the relativistic mean field theory as explained in the numerical
calculation section.

In order to incorporate
higher order correction by the $\pi$ and $\rho$ exchange interaction,
we carry out the random phase approximation which 
sums up ring diagrams to all orders and is expressed by Dyson's
equation 
\begin{equation}
\label{eq:RPA}
\Pi^{\rm RPA}=\Pi^0+\Pi^0 V\Pi^{\rm RPA}.
\end{equation}
$\Pi^0$ is the free response function defined above and $\Pi^{\rm
RPA}$ is the full response function obtained by RPA. 
$V$ is $\pi$ and $\rho$ meson exchange interaction,
\begin{eqnarray}
\label{eq:piex}
V_\pi (q)&=&\left (\frac{f_\pi}{m_\pi}\right
)^2\left ( \frac{1}{q^2-m_\pi^2}-\frac{g'_\pi}{q^2}\right )\\ 
\label{eq:rhoex}
V_\rho^{\mu\nu}(q)&=&g_\rho^2(g^{\mu\nu}-q^\mu q^\nu/m_\rho^2)
\left (\frac{1}{q^2-m_\rho^2}-C_\rho^{-1}\frac{g'_\rho}{q^2}\right ).
\end{eqnarray}
$C_\rho=2.18$ is the ratio of couplings of non-relativistic
interactions 
$\left
(\frac{f_\rho}{m_\rho}\right )^2/\left (\frac{f_\pi}{m_\pi}\right )^2$ 
and $g'_\pi$ and $g'_\rho$ are 'relativistic' phenomenological
Landau-Migdal parameters introduced to take into account short range
nuclear correlations. In reality the RPA equation for the
polarization function induced by the $\rho$ exchange involves Lorentz
contraction. We can, however, deduce the RPA equation to
the same form as that for $R_L$ by
consideration that
$q^\mu\Pi_{\mu\nu}=0$ and decomposition,
$\Pi_{\mu\nu}=P^L_{\mu\nu}\Pi_L^{(\rho)} +P^T_{\mu\nu}\Pi_T^{(\rho)}$.
Here $P^{L,T}_{\mu\nu}$ are projection operators which extract the
longitudinal 
and transverse part $\Pi_{L,T}^{(\rho)}$ of the $\rho$-induced
response\cite{Hatsuda}. Finally we obtain
the RPA response functions in a simple form, 
\begin{eqnarray}
\Pi_L^{\rm RPA}&=&(1-V_\pi\Pi_L^0)^{-1}\Pi_L^0\\
\Pi_T^{\rm RPA}&=&(1-V_\rho\Pi_T^0)^{-1}\Pi_T^0
\end{eqnarray}
with 
\begin{equation}
V_\rho(q)=g_\rho^2\left
(\frac{1}{q^2-m_\rho^2}-C_\rho^{-1}\frac{g'_\rho}{q^2}\right ). 
\end{equation}
We adopt $g_\rho=2.6$ which satisfies 
\begin{equation}
\frac{f_\rho}{m_\rho}\approx g_\rho\frac{\kappa +1}{2M},
\end{equation}
in accordance with the non-relativistic coupling constant.

The Landau-Migdal parameter was originally introduced in the
non-relativistic $\pi\!+\!\rho\!+\!g'$ model and in that case
the non-relativistic Landau-Migdal parameter is
$g'=g'_\pi\!=\!g'_\rho$,  since 
\begin{eqnarray}\nonumber
\label{Eq:constraint}
V_{g'}&=&g'\left ( \frac{f_\pi}{m_\pi}\right )^2\bold{\sigma}_1\!\cdot
\!\bold{\sigma}_2\,\bold{\tau}_1\!\cdot\!\bold{\tau}_2\\
&=&g'\left \{ \left ( \frac{f_\pi}{m_\pi}\right
)^2\bold{\sigma}_1\!\cdot \!\hat{q}\, \bold{\sigma}_2\!\cdot
\!\hat{q}+\left 
( \frac{f_\rho}{m_\rho}\right )^2C_\rho^{-1} 
\bold{\sigma}_1\!\times \!\hat{q}\cdot \bold{\sigma}_2\!\times
\!\hat{q}\right  
\}\bold{\tau}_1\!\cdot\!\bold{\tau}_2.
\end{eqnarray}
$g'=0.6\sim 0.7$ is usually used as a standard value.
On the other hand, the longitudinal correlation $g'_\pi$ and the
transverse correlation $g'_\rho$ defined in Eqs.(\ref{eq:piex}) and
(\ref{eq:rhoex}) are in general different.
It is very interesting to extract these informations from the G-matrix 
of the relativistic Br\"{u}ckner-Hartree-Fock
theory\cite{Fuchs}. 
Here, we extract them from inelastic electron scattering experiments.
Horowitz et.al. studied the transverse response involved with $(e,e')$ 
scattering\cite{Horowitz}, from which they obtained $g'_\rho=0.3$ to
reproduce the data. As for $g'_\pi$, we choose the conventional value, 
$g'_\pi=0.7$.
Thus  
we employ here the parameter set
$g'_\rho=0.3,g'_\pi=0.7$.

In (p,n) reactions the distortion effect is of much importance compared
with $(e,e')$ reactions due to large nucleon-nucleon cross section,
$\sigma_{NN}$. We carry out the local density approximation with
the attenuation factor given by the eikonal approximation to obtain 
the $(\vec{p},\vec{n})$ cross sections by using the nuclear matter
calculation. 
The response functions for finite nucleus are expressed by
\begin{equation}
\label{Eq:LDA}
R^{\rm LDA}_{T,L}=\frac{1}{N_{\rm eff}}\frac{N}{A}\int 2\pi b \,db \,dz
e^{-\chi(b)}\rho(b,z)R_{T,L}(k_F) .
\end{equation}
$e^{-\chi(b)}$ is the attenuation factor for the reaction with the
impact parameter $b$ where
$\chi(b)=\int_{-\infty}^{+\infty}dz\sigma_{NN}\rho(b,z)$. 
$\rho(b,z)$ is the nuclear density profile of the target nucleus in the
cylindrical coordinate. The effective neutron number $N_{\rm eff}$ is
given as the normalization factor 
\begin{eqnarray}\nonumber
N_{\rm eff}&=&\frac{N}{A}{\int 2\pi b \,db\, dz
e^{-\chi(b)}\rho(b,z)}\\ \label{Eq:Neffective}
&=&\frac{N}{A}\frac{1}{\sigma_{NN}}\int 2\pi b \,db \,\chi(b)
e^{-\chi(b)}. 
\end{eqnarray}
Here the factor $\frac{N}{A}$ appears for the use of the neutron
density instead of nucleon density in
Eq.(\ref{Eq:LDA}),(\ref{Eq:Neffective}). 
For the nuclear density distribution $\rho(\bold{r})$ we adopt the
three-parameter 
Fermi-type form determined experimentally~\cite{Table}. The response
function $R(k_F)$ is obtained by the nuclear matter calculation using
the Fermi momentum and the 
effective mass given by 
the relativistic mean field calculation for nuclear
matter~\cite{Hirata}, at the corresponding density.
We use $\sigma_{NN}\!=\!30$mb which is close to the free cross section
at the relevant energy $T_{\rm lab}\approx 400$MeV, since the reaction 
is supposed to take place close to the nuclear surface.
The typical Fermi momentum $\bar{k}_F$ is obtained by
replacing $R(\rho(b,z))$ in Eq.(\ref{Eq:LDA}) by $k_F\sim \rho^{1/3}$.
We find $\bar{k}_F= 0.98(0.97)$ for
$^{12}$C($^{40}$Ca) with  $M^{*}$ at this density being
$M^{*}=0.82(0.83)M$.

\section{Numerical results} 
We show the numerical results on the response functions at 
$\mid\!\bold{q}\!\mid=\!1.7$fm$^{-1}$ which are of special interest
here and measured by the experiments.
To begin with we demonstrate relativistic effects on free response
functions of nuclear matter at different densities.
In Fig.1 we show the free response functions for the
symmetric nuclear matter at the
saturation density $k_F\!=\!1.36$fm$^{-1}$, with the effective mass
$M^*\!=\!0.59M$  
which is given by the RMF calculation. For comparison the
non-relativistic response function and the relativistic response with
$M^{*}\!=\!M$ are also shown in the figure. 
We see that the longitudinal spin response $R_L$ with $M^*\!=\!0.59M$ is
reduced from 
the non-relativistic response due to the effect of the effective mass. 
This is
affirmed by seeing that $R_L$ becomes close to $R_{\rm NR}$ by setting 
$M^*\!=\!M$, which are, as stated in the previous section, identical
with $R_{\rm NR}$ 
for $\omega <30$MeV. 
Broadening of the width is a consequence of enlarged maximum
particle-hole excitation energy for given momentum transfer
$\sqrt{(k_F+\mid\! 
\bold{q}\!\mid)^2+M^{*2}}-\sqrt{k_F^2+M^{*2}}\sim 
\frac{k_F\mid \bold{q}\mid}{M^{*}}$.
We find the transverse response $R_T$ shows more moderate effective
mass dependence, and it is larger than the longitudinal response with
the same effective mass. 
We mention here that the effective mass also makes the
non-relativistic responses smaller as the case of the relativistic
responses\cite{Futami,Meyer-ter-Vehn,Shiino}. There are, however, two
major 
differences between the 
relativistic and the non-relativistic cases. First, the effective mass 
is much smaller for the relativistic case. Second, while the effective
mass effect works in the same way for both the spin longitudinal and
the transverse responses in the non-relativistic case, 
in the relativistic 
case, the longitudinal response is more reduced than the 
transverse one as shown here and in the previous section.
As a result we find the ratio
$R_L /R_T\approx 0.8$ with little dependence on $\omega$.
 
In Fig.2 the free response functions for
$k_F=0.97$fm$^{-1}$($M^*\!=\!0.83$) 
are shown, which corresponds to the density $\rho=0.37\rho_0$. 
$(\vec{p},\vec{n})$ reaction is supposed to take place in the surface
region of a nucleus with about this density because of the large
distortion of the incident and outgoing nucleon. 
The width becomes narrower and the magnitude becomes larger than those
at the saturation density due to the small Fermi momentum and
the larger effective mass. The qualitative features in the previous
figure hold also true here. 
The difference between $R_L$ and $R_T$ is also smaller than Fig.1,
which yields  $R_L/R_T \approx 0.94$. We find that
in any case $R_L/R_T <1$ holds in the relativistic case. This fact is
crucial to reproduce the experimental observation $R_L/R_T\approx 1$.

Next we show the relativistic responses calculated by RPA.
In Fig.3(a) we show the longitudinal responses with RPA at
the saturation density together with free response and compare with
the non-relativistic results. 
We see that the relativistic response with RPA correlation
, $R_L^{\rm RPA}$, is 
only slightly enhanced by the attractive $\pi$ exchange interaction
while the non-relativistic one is largely enhanced.
The difference from the
non-relativistic case originates from the effective coupling reduced
by the 
factor $\left ( \frac{M^*}{M}\right )^2$ which makes $R_L^{\rm RPA}$
closer to the free response, $R_L^{\rm RPA}\sim R_L^{\rm Free}$.
In Fig.3(b), on the other hand, we show the transverse responses at
the same density. 
It is found that
the relativistic response with RPA, $R_T^{\rm RPA}$,
almost agrees with 
the free response. This arises from  $V_\rho\approx 0$ with
$g'_\rho\!=\!0.3$, 
while the non-relativistic response is considerably quenched by
the repulsive 
$\rho$ exchange force with $g'\!=\!0.7$.
We find that the reduction of the relativistic longitudinal response
compared with 
the transverse response is compensated by the RPA correlation.
It follows that $R_L^{\rm RPA}/R_T^{\rm RPA}\approx 1$ at low
$\omega$(Fig.5). 
It is in good contrast with the non-relativistic case where
significant quenching of the transverse response and enhancement of
the longitudinal one is found to yield $R_L^{\rm RPA}/R_T^{\rm RPA}
\gg 1$.

We note here that the use of the smaller 'relativistic' $g'_{\rho}$ in 
the transverse channel is supported by the similarity of the
transverse responses, $R_T^{\rm RPA}$, shown by the solid curve and by 
the thin dotted line curve in the smaller $\omega$ region in
Fig.3(b). The relativistic free response is already reduced largely
from the non-relativistic one.

In Fig.4(a), (b) we show the responses with RPA at
$k_F\!=\!0.97$fm$^{-1}$. 
The dependence of $R^{\rm free}$ on $k_F$ causes the attractive
$\pi$ exchange force weaker in both the non-relativistic and relativistic
cases. In the latter case the factor $\frac{M^*}{M}$ reduces still more 
the effective coupling, hence less enhancement by RPA correlation.
The relativistic transverse response $R_T^{\rm RPA}$
agrees with the free response for the same reason as in the case of
the saturation density. 
We obtain the ratio 
$R_L^{\rm RPA}/R_T^{\rm RPA}=1\sim 1.2$ at low $\omega$, while the
non-relativistic case 
results in $R_L^{\rm RPA}/R_T^{\rm RPA} >2$ with striking $\omega$
dependence as shown in Fig.5. 
The relativistic results are in agreement with the
experimental situation.

Finally we compare the results obtained by the local density
approximation with the experimental data recently measured at
RCNP~\cite{Wakasa}. 
Fig.6 shows the results 
on $^{12}$C(the left panels) and $^{40}$Ca(the right panels).
The in-medium cross section $\sigma_{NN}\!=\!30$mb provides
effective neutron number 
$N_{\rm eff}\!=\!2.0(3.7)$ for
$^{12}$C($^{40}$Ca).  
Since our effective neutron number differs from the
experimental one, we multiply $N_{\rm eff}^{\rm exp}/N_{\rm
eff}$ to the experimental data 
where $N_{\rm eff}^{\rm exp}$ is
the effective neutron number adopted in the experimental analysis,
$N_{\rm eff}^{\rm exp}=2.7$(6.0) for $^{12}$C and
$^{40}$Ca, respectively. 
This is because the ``experimental'' response functions extracted from
the cross sections and the polarization measurements by the
experimentalists were obtained by dividing the corresponding
quantities with $N_{\rm eff}^{\rm exp}$\cite{LAMPF,Wakasa}. On the other
hand, the ``theoretical'' response functions are obtained by dividing
the corresponding quantities with $N_{\rm eff}$ as shown in
Eq.(\ref{Eq:LDA}). 
As seen in the discussion above about the matter properties, the
longitudinal and the transverse responses are very close to each
other. Below $\omega <60$MeV, the theoretical results are close to the 
experimental responses (The use of smaller in-medium cross section
$\sigma_{NN}$ 
makes the agreement better in $^{12}$C). However, the theoretical
responses are much 
smaller than the experimental results above $\omega >60$MeV. 
For the full understanding of the problem of the spin response
functions, we have to describe this deviation. The $\Delta$ excitation 
starts to contribute above $\omega >80$MeV as can be seen from the
$(e,e')$ data\cite{Horowitz}. It would be very interesting to work out
the two and multiple scattering processes in the $(p,p')$ and $(p,n)$
reactions for the spin responses\cite{Pace}.
\section{Conclusion}
We have studied the relativistic effects on the spin response functions 
by using the relativistic version of the $\pi\!+\!\rho\!+\!g'$ model.
We have calculated the polarization functions in the $\pi$(spin
longitudinal) channel and $\rho$ meson(spin transverse) channel in the
nuclear medium to obtain the spin response functions. 
We have adopted the pseudovector coupling for $\pi NN$ channel instead
of the pseudoscalar coupling. The longitudinal
response calculated with the pseudovector coupling is largely
reduced by the effective mass which is 
$\frac{M^*}{M}=0.6\sim 0.8$ in the nuclear medium. 
The reduction factor is found to be $\left (\frac{M^*}{M}\right )^2$
at the higher density  
and $\frac{M^*}{M}$ at the lower density.
On the other hand the transverse response is reduced by
$\left ( 1+\kappa\frac{M^*}{M}\right )^2/(1+\kappa)^2$ at the high
density 
because of the vector-tensor coupling for $\rho NN$ vertex.
We have found that the
relativistic longitudinal and transverse 
responses agree with the non-relativistic ones in the case of $M^*=M$.
We have shown that without RPA correlation the longitudinal response is
less than the transverse response even at the low density
because of the different effective mass dependence.

The reduction of the effective mass also reduces the higher order 
$\pi NN$ and $\rho NN$ correlations in the medium. We have demonstrated
that the relativistic longitudinal response function is enhanced much
less than the non-relativistic case by the attractive $\pi$ exchange
force. The transverse response with RPA correlation has been
calculated with the 
Landau-Migdal parameter $g'_\rho=0.3$, which is chosen by Horowitz
{\it et al}. as a fitting parameter in the analysis of electron
scattering. 
We have shown that the transverse response is hardly affected by RPA.
As a result we have obtained similar response functions
for the longitudinal and the transverse spin responses. It is in
agreement with the experimental data at least in the small $\omega$
region, which is qualitatively different from the
non-relativistic theoretical results.

We have calculated the response functions for $^{12}$C and $^{40}$Ca
by using the eikonal prescription for the distortion effect with the
use of the response functions in nuclear matter. We have compared 
the longitudinal and the transverse response functions directly with
the recently obtained experimental results. We see close reproductions 
of the spin responses in the small $\omega$ region($\omega<60$MeV) for 
both the longitudinal and the transverse channels. The result is due
largely to the use of the relativistic responses with the pseudovector 
coupling in the $\pi$ channel with large reduction of the effective
nucleon mass in the nuclear medium. The use of the smaller
$g'_\rho$($g'_\rho=0.3$) is adopted for another reason. This small
value is extracted from $(e,e')$ reactions phenomenologically and
needs to be studied theoretically in the relativistic G-matrix
theory. 

The theoretical responses underestimate the experimental ones largely
above $\omega>60$MeV. We expect two and more step processes at this
high excitation energies, which should be worked out
quantitatively. We start to see the effect of the $\Delta$ isobar
excitations already starting around $\omega >80$MeV, which are seen
in the $(e,e')$ reaction data. Unless these effects are quantitatively 
studied, our theoretical results discussed in this paper is considered 
as a plausible explanation of the long standing puzzle on the ratios
of the longitudinal and the transverse responses.
\section*{Acknowledgment}
We are grateful to H. Sakai, T. Wakasa, M. Fujiwara and I. Tanihata
for fruitful discussions on the experimental aspects of (p,n)
reactions.

\newpage
\begin{figure}[htbp]
\label{Fig:Free-norm}
\centerline{
\epsfxsize=100mm\epsffile{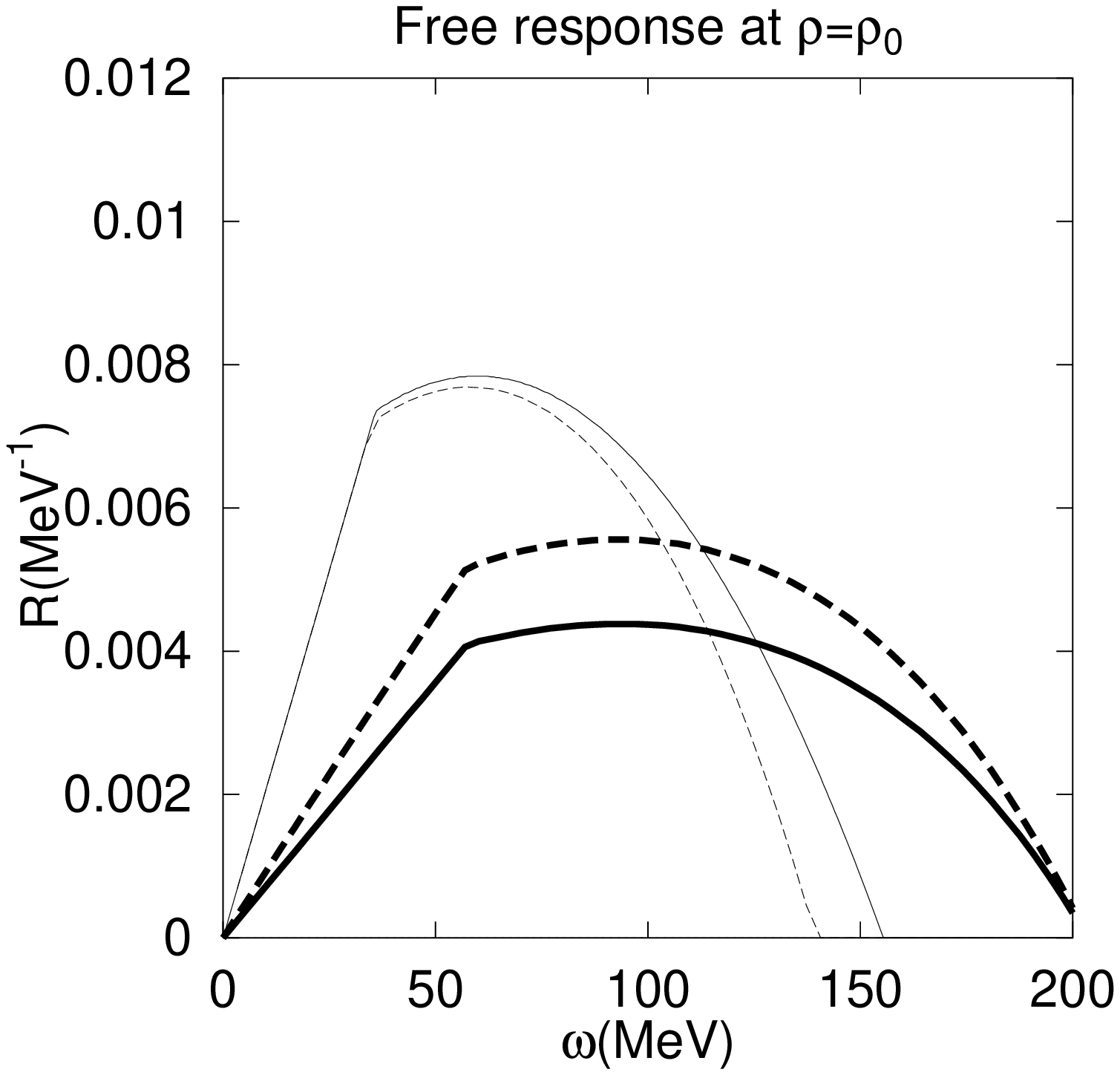}}
\caption{ Free relativistic longitudinal response function $R_L$
(thick solid curve)  and  
transverse response
function $R_T$ (thick dashed) and non-relativistic one $R_{\rm NR}$
(thin solid) at $\mid\!\!\bold{q}\!\!\mid=\!1.7$fm$^{-1}$ for the
symmetric nuclear  
matter as functions of the excitation energy $\omega$ at the
saturation density with the Fermi momentum,
$k_F\!=\!1.36$fm$^{-1}$. For comparison the relativistic 
longitudinal response with $M^{*}\!=\!M$ is shown(thin dashed).
}
\end{figure}
\begin{figure}[htbp]
\label{Fig:Free-ref}
\centerline{
\epsfxsize=100mm\epsffile{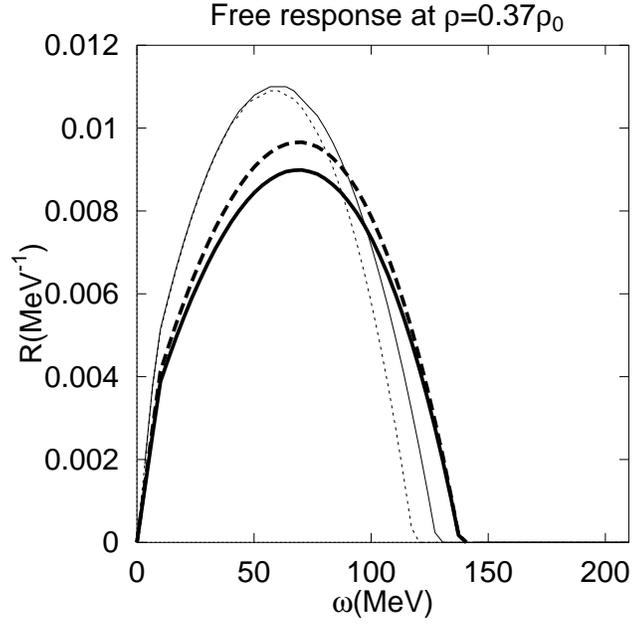}}
\caption{ Same as  Fig.1 for
$k_F=0.97$fm$^{-1}$ which corresponds to the typical density of the
nucleus in  
$(p,n)$ reaction
because of the large distortion effects on the incident and outgoing
nucleons.} 
\end{figure}
\begin{figure}[htbp]
\centerline{
\epsfxsize=75mm\epsffile{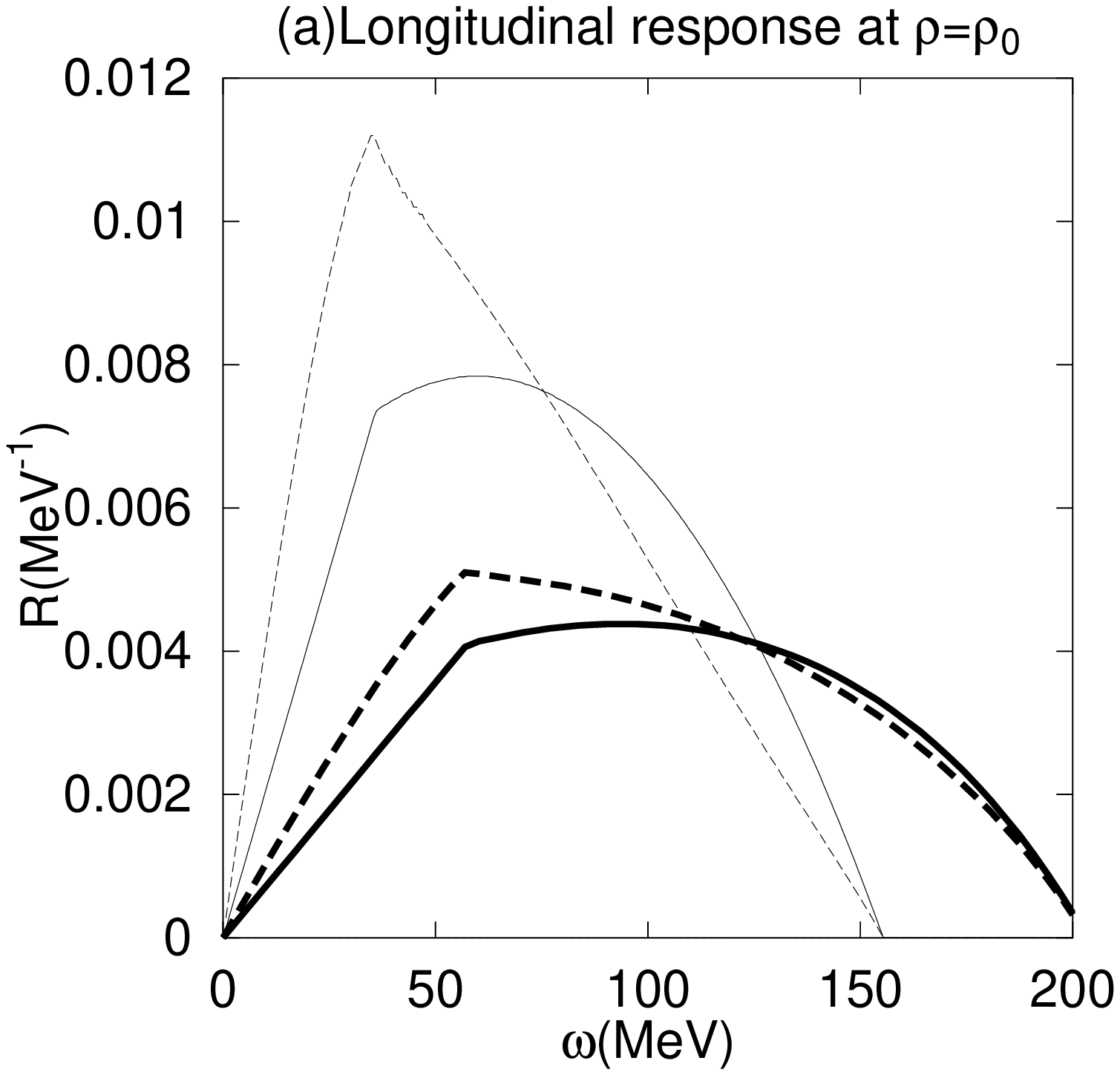}
\epsfxsize=75mm\epsffile{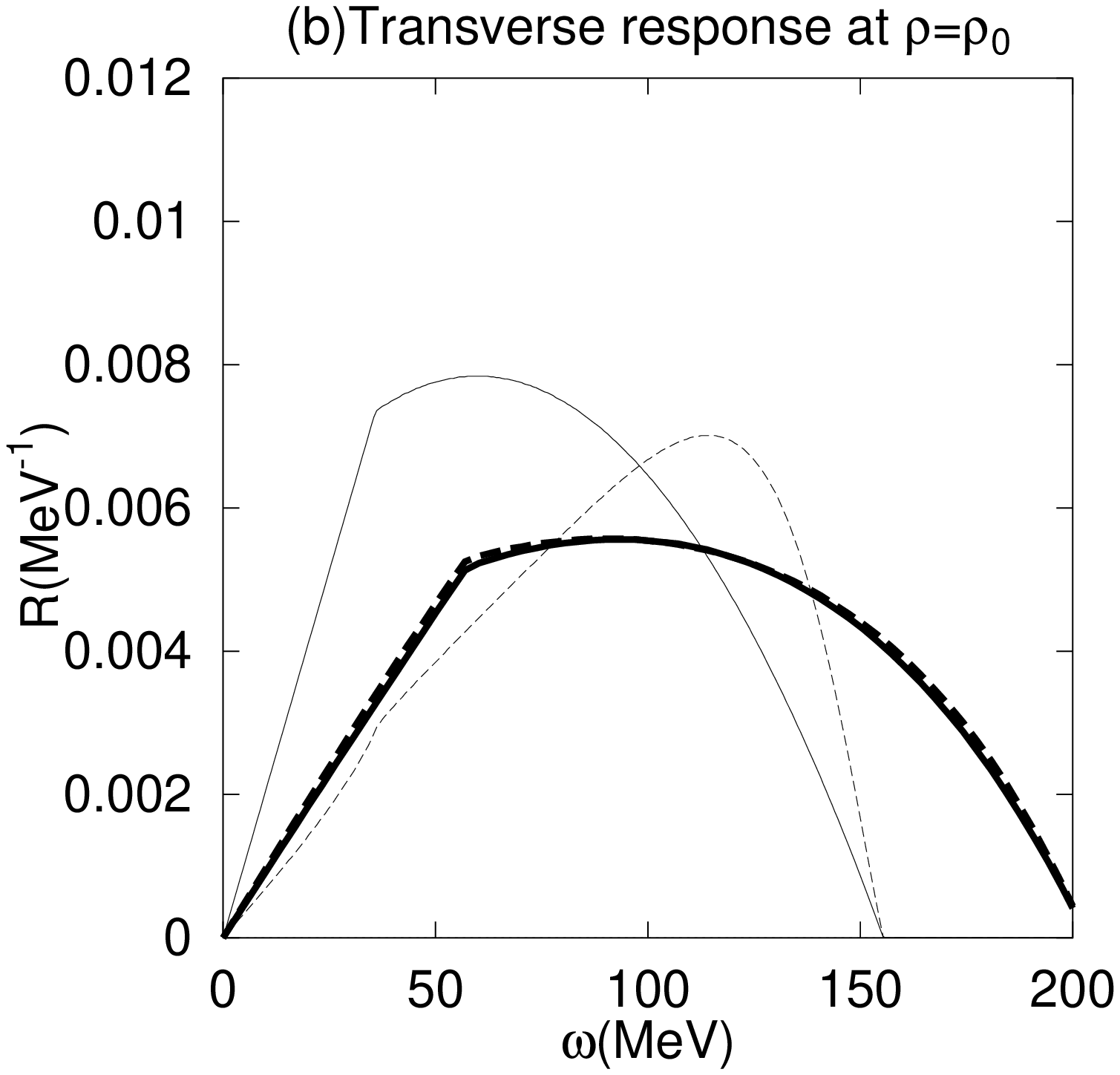}}
\caption{ (a)Longitudinal  and (b)transverse
responses with and without RPA
correlation at the saturation 
density,$k_F=1.36$fm$^{-1}$.
Thick solid curves indicate the relativistic free
responses $R_L^{\rm free}$ and $R_T^{\rm free}$ and thick dashed curves
RPA results $R_L^{\rm RPA}$ and $R_T^{\rm RPA}$. The corresponding
non-relativistic values  
are drawn with thin solid and dashed curves.
}
\end{figure}
\begin{figure}[htbp]
\centerline{
\epsfxsize=75mm\epsffile{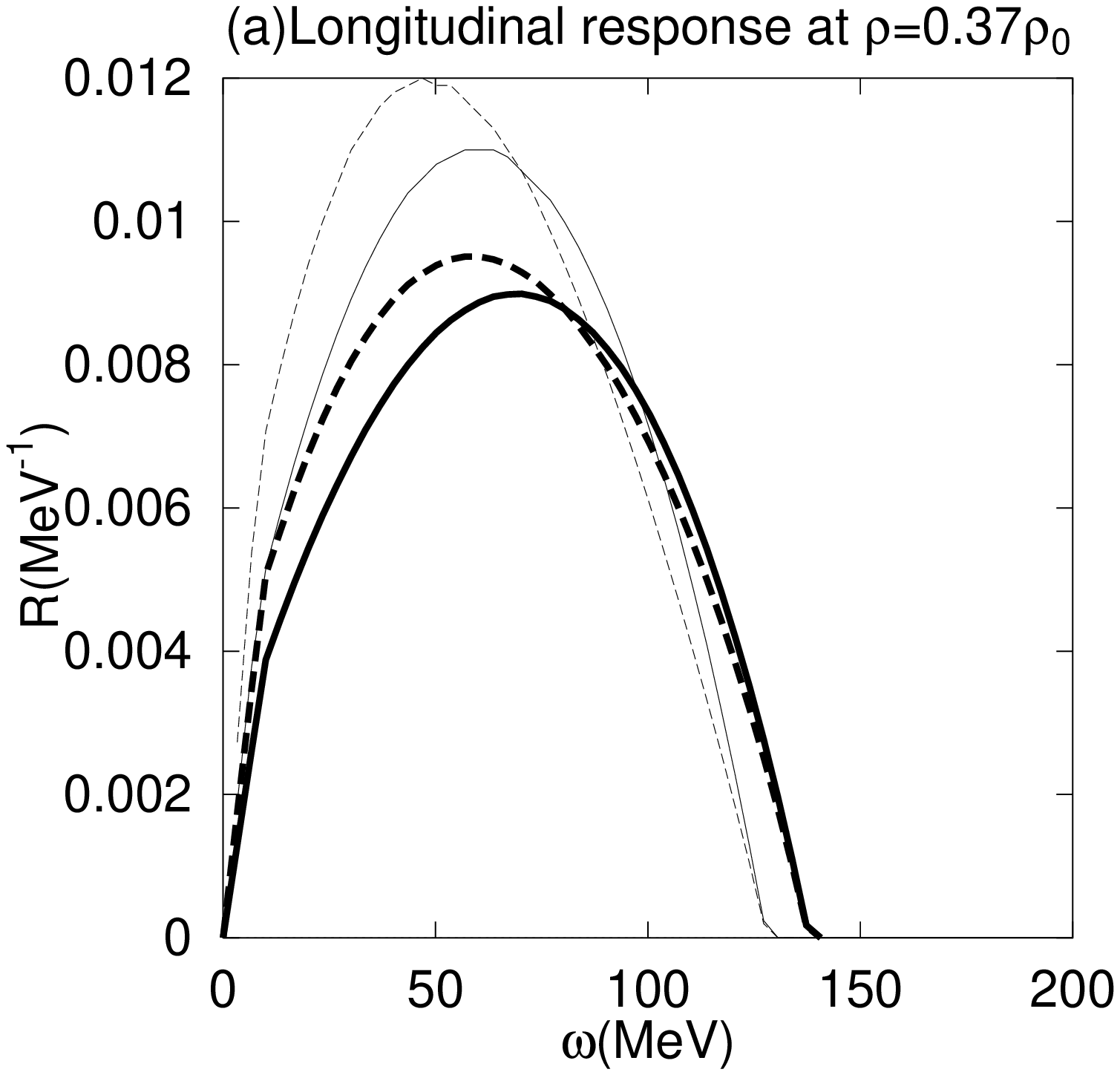}
\epsfxsize=75mm\epsffile{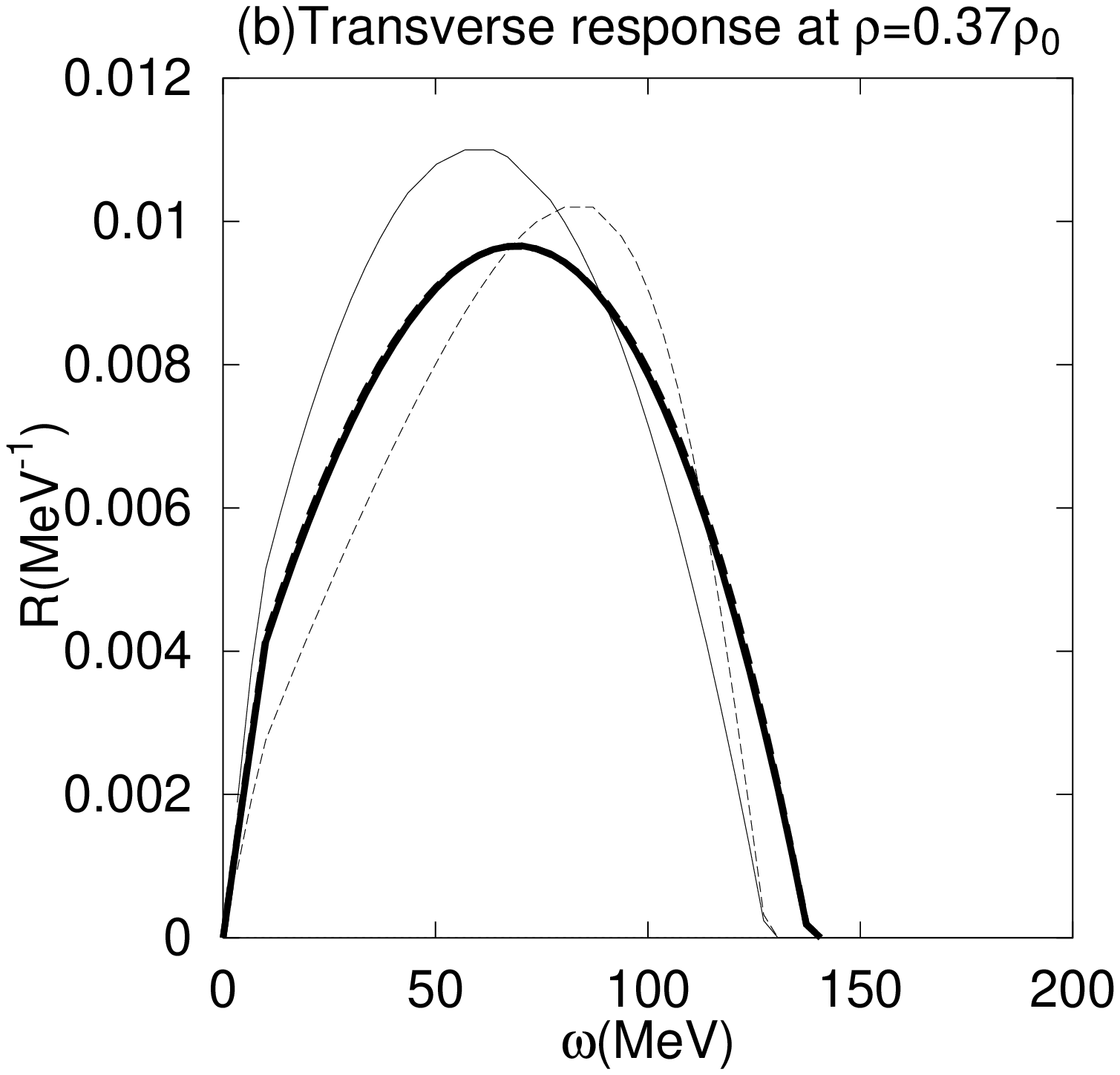}}
\caption{ Same as Fig.3 for $k_F=0.97$fm$^{-1}$.
}
\end{figure}
\begin{figure}[htbp]
\centerline{
\epsfxsize=100mm\epsffile{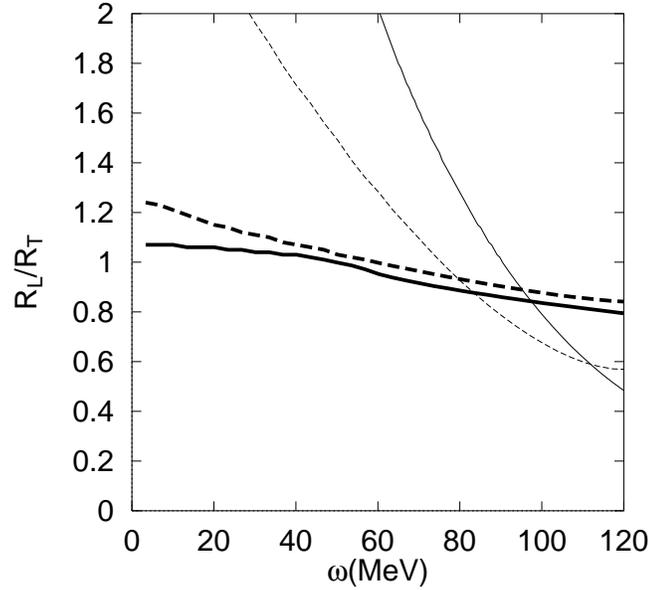}}
\caption{ Ratio of the longitudinal response to the transverse one
with 
RPA at the saturation density and $k_F=0.97$fm$^{-1}$corresponding to
$\rho=0.37\rho_0$. A thick 
solid curve indicates the relativistic result at the saturation
density and  
a thick dashed curve for
$k_F=0.97$fm$^{-1}$. Thin solid and dashed curves indicate the
non-relativistic results at the saturation density and at
$\rho=0.37\rho_0$, respectively.
}
\end{figure}
\begin{figure}[htbp]
\begin{minipage}[tbp]{70mm}
\rightline{
\epsfxsize=75mm\epsffile{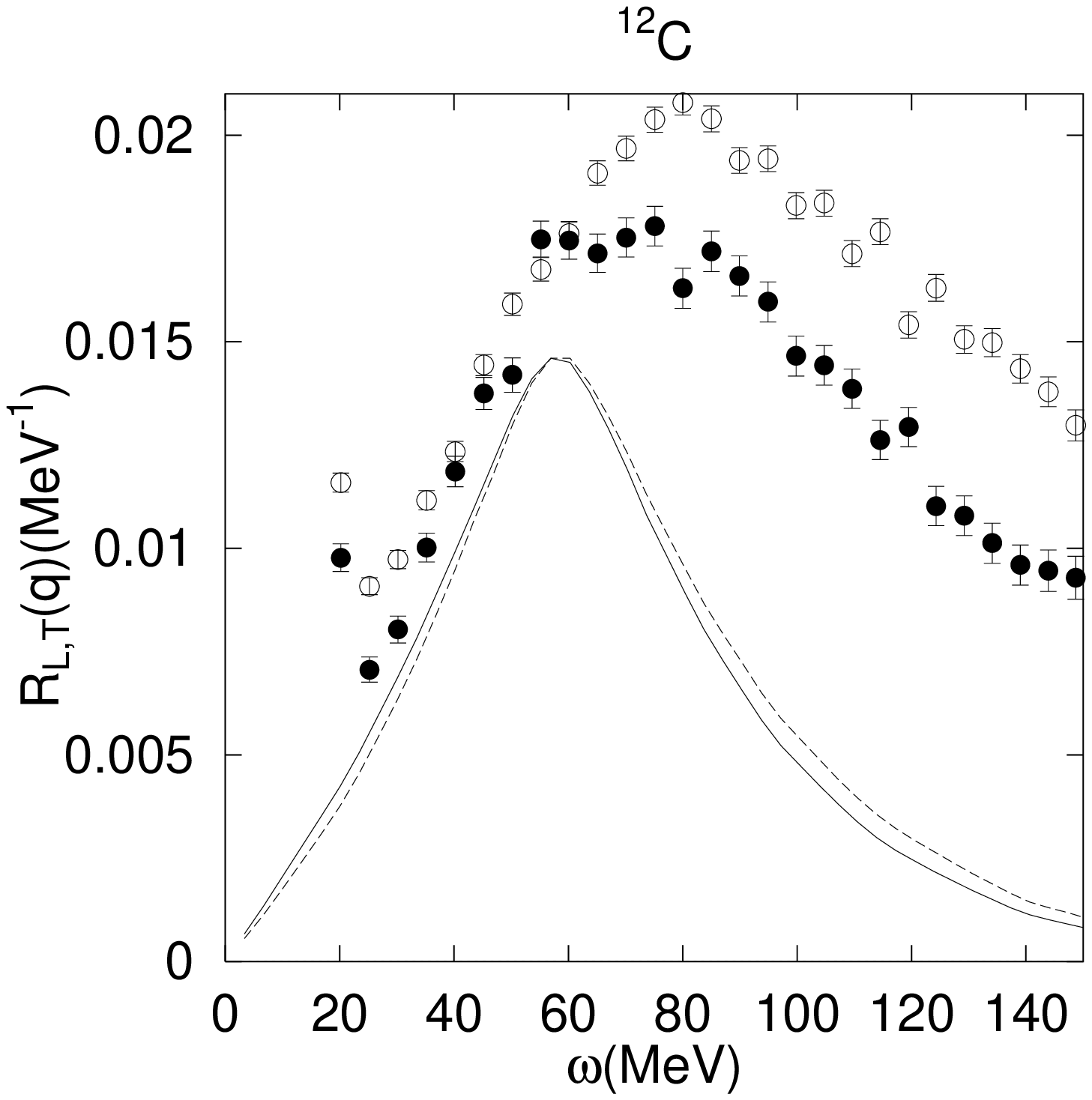}}
\rightline{
\epsfxsize=75mm\epsffile{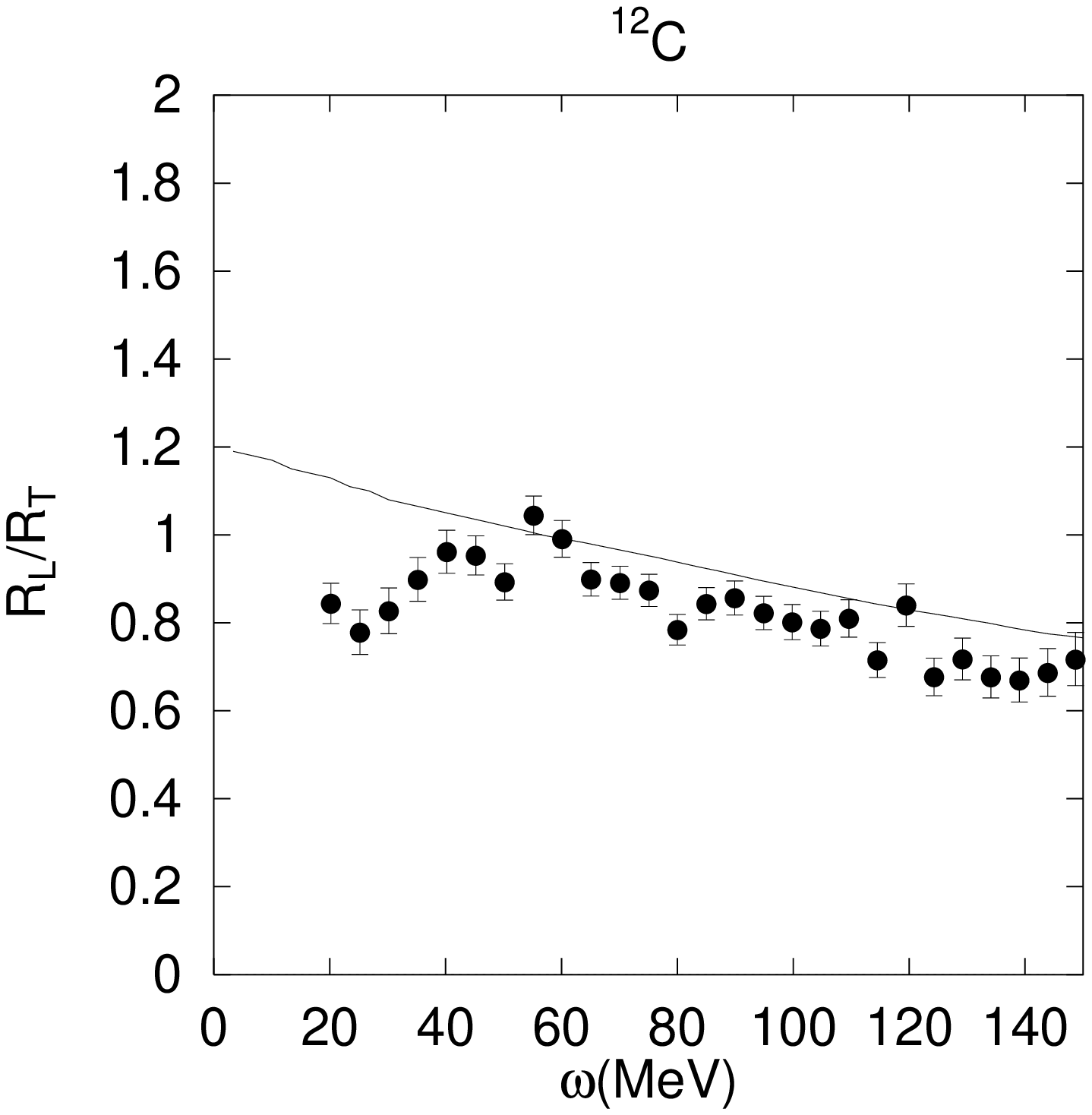}}
\end{minipage}
\hskip 5mm
\begin{minipage}[t]{70mm}
\rightline{
\epsfxsize=75mm\epsffile{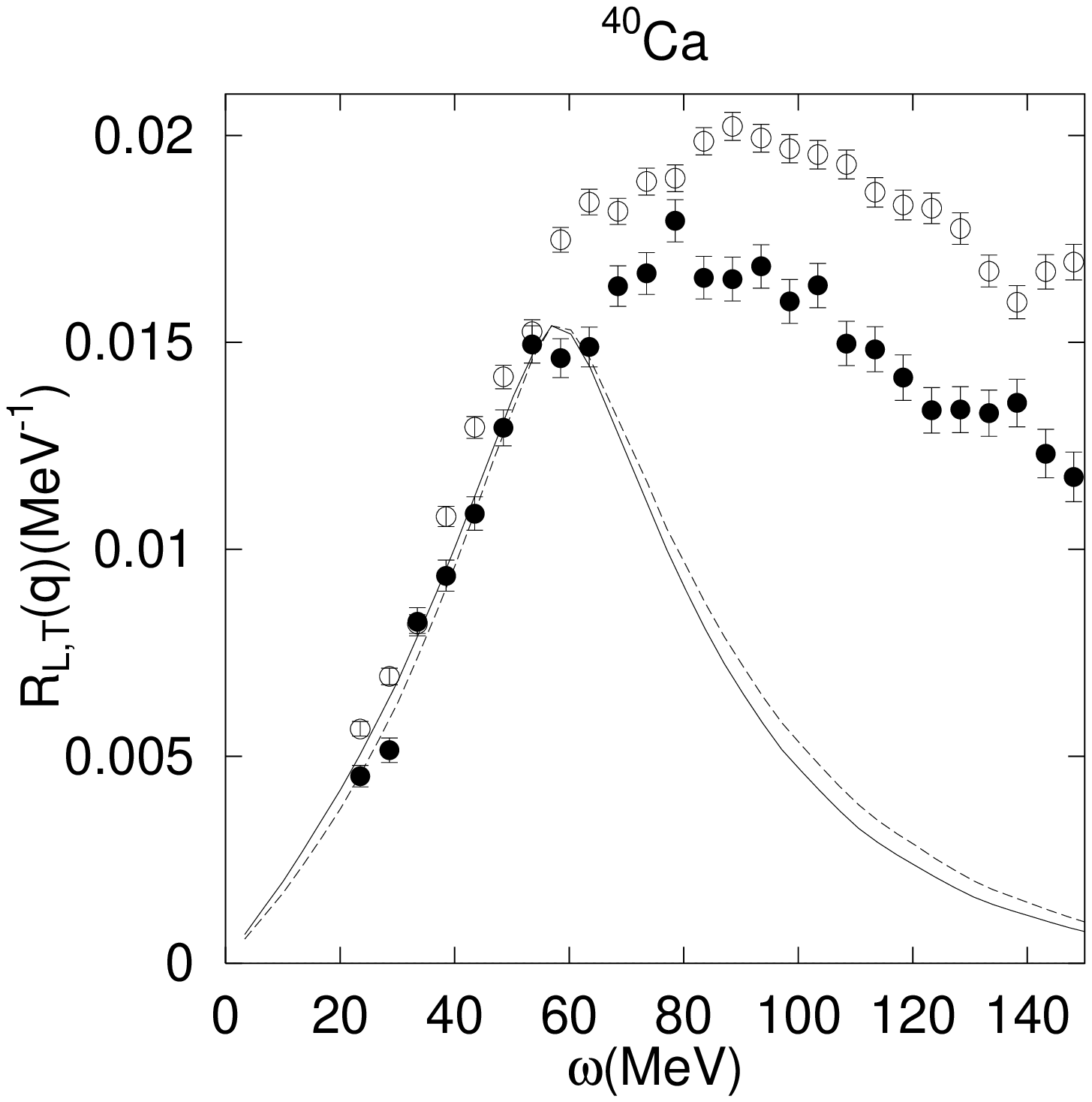}}
\rightline{
\epsfxsize=75mm\epsffile{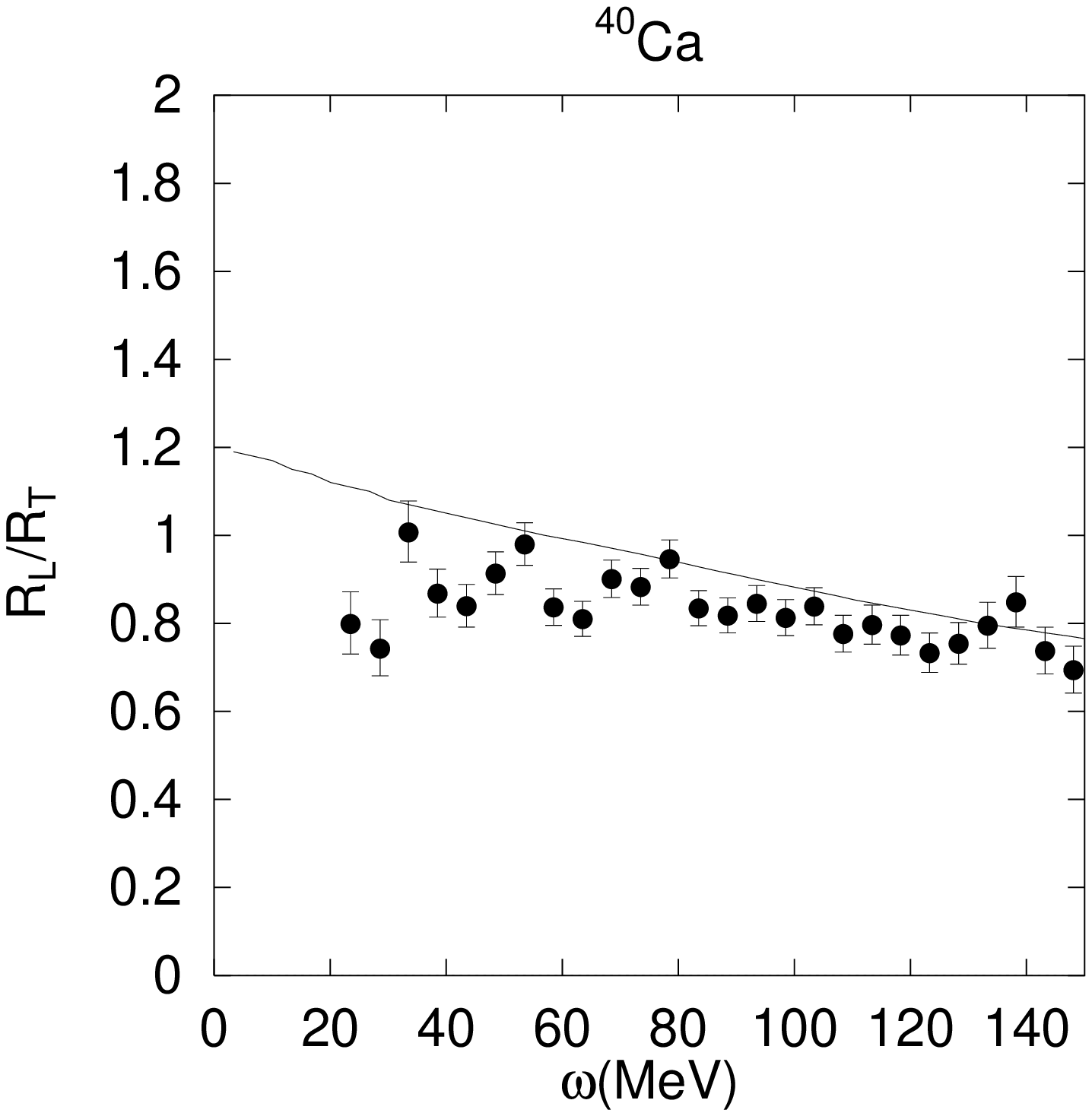}}
\end{minipage}
\caption{ Response functions with RPA(upper panels) and
ratio of the 
longitudinal to the transverse response(lower panels) at
$\mid\!\bold{q}\!\mid= 1.7$fm$^{-1}$
obtained by the local density
approximation  with $\sigma_{NN}=30$mb for $^{12}$C(left panels) and
$^{40}$Ca(right panels).
In the upper panels the longitudinal response is drawn by a solid
curve and the transverse by a dashed curve.
The experimental data are plotted by solid circles for the longitudinal 
response and by open circles for the transverse one. The experimental
responses are 
multiplied by a factor $N_{\rm eff}^{\rm exp}/N_{\rm eff}$(see 
the text).
In the lower panels
the theoretical value is drawn by a solid curve and the experimental
data are plotted by solid circles with error bars.
}
\end{figure}

\begin{thebibliography}{99}
\bibitem{Migdal} A.B. Migdal, Rev. Mod. Phys. {\bf 50}(1978)107 and
references therein. 
\bibitem{Toki} H. Toki and W. Weise, Phys. Rev. Lett. {\bf 42}(1979)1034.
\bibitem{Ericson} W.M. Alberico, M. Ericson and A. Molinari,
Nucl. Phys. {\bf A379}(1982)429.
\bibitem{LAMPF}T.A. Carey, K.W. Jones, J.B. McClelland, J.M. Moss,
L.B. Rees, N. Tanaka and A.D. Bacher, Phys. Rev. Lett. {\bf
53}(1984)144.\\ 
L.B. Rees, J.M. Moss, T.A. Carey, K.W. Jones, J.B. McClelland, N. Tanaka,
A.D. Bacher and H. Esbensen, Phys. Rev. {\bf C34}(1986)627.\\
J.B. McClelland {\it et al}., Phys. Rev. Lett. {\bf 69}(1992)582.\\
X.Y. Chen et.al, Phys. Rev. {\bf C47}(1993)2159.\\
T.N. Taddeucci {\it et al}., Phys. Rev. Lett. {\bf 73}(1994)3516.
\bibitem{RCNP} H. Sakai, M.B. Greenfield, K. Hatanaka, S. Ishida,
N. Koori, H. Okamura, A. Okihana, H. Otsu, N. Sakamoto, Y. Satou,
T. Uesaka and T. Wakasa, Nucl. Phys. {\bf A577}(1994)111c.
\bibitem{Wakasa} 
T. Wakasa {\it et al}., submitted to Phys. Rev. {\bf C}.\\
T. Wakasa, private communication.
\bibitem{Esbensen} H. Esbensen, H. Toki and G. Bertsch, Phys. Rev. {\bf
C31}(1985)1816.
\bibitem{Ichimura} M. Ichimura, K. Kawahigashi, T.S. J\o rgensen and
C. Gaarde, Phys. Rev. {\bf C39}(1989)1446.
\bibitem{Savushkin} B.L. Birbrair, V.N. Fomenko and L.N. Savushkin,
J. Phys. {\bf G8}(1982)1517. 
\bibitem{Horowitz} C.J. Horowitz and J. Piekarewicz,
Phys.Lett. {\bf B301}(1993)321.\\
C.J. Horowitz and J. Piekarewicz, Phys.Rev. {\bf C50}(1994)2540.
\bibitem{Walecka} J.D. Walecka, Ann. Phys. {\bf 83}(1974)491.
\bibitem{Serot} B.D. Serot and J.D. Walecka, Adv. in Nucl.Phys. {\bf
16}(1986).
\bibitem{Brockmann} R. Brockmann and R. Machleidt, Phys. Rev. {\bf
C42}(1990)1965.
\bibitem{Hirata} D. Hirata ,H. Toki, T. Watabe, I. Tanihata and
B.V. Carlson, Phys. Rev. {\bf C44}(1991)1467.
\bibitem{Sugahara} Y. Sugahara and H. Toki, Nucl. Phys. {\bf
A579}(1994)557. 
\bibitem{Oset} E. Oset, H. Toki and W. Weise, Phys. Rep. {\bf
83}(1982)281.
\bibitem{Hohler} G. H\"{o}hler and E. Pietarinen, Nucl. Phys. {\bf
B95}(1975)210. 
\bibitem{Matsui} T. Matsui and B.D. Serot, Ann. Phys. {\bf 144}(1982)107.
\bibitem{Hatsuda}H. Shiomi and T. Hatsuda, Phys.Lett. {\bf B334}(1994)281.
\bibitem{Dawson} J.F. Dawson and J. Piekarewicz, Phys. Rev. {\bf
C43}(1991)2631.
\bibitem{Kurasawa} H. Kurasawa and T. Suzuki,
Nucl. Phys. {\bf A445}(1985)685.
\bibitem{Walecka-text}A.L. Fetter and J.D. Walecka, Quantum theory of
many-particle systems (McGraw-Hill, New York, 1971).
\bibitem{Fuchs} C. Fuchs, L. Sehn and H.H. Wolter, Nucl. Phys. {\bf
A601}(1996)505. 
\bibitem{Table} C.W. de Jager, H. de Vries and C. de Vries, At. Data
Nucl. Data Tables {\bf 14}(1974)479.
\bibitem{Futami} H. Toki, Y. Futami and W. Weise, Phys. Lett. {\bf
B78}(1978)547. 
\bibitem{Meyer-ter-Vehn}W.H. Dickhoff, A. Faessler, H. Muther and 
J. Meyer-Ter-Vehn, Nucl. Phys. {\bf A368}(1981)445.\\
W.H. Dickhoff, A. Faessler, J. Meyer-Ter-Vehn and  H. Muther,
Phys. Rev. {\bf C23}(1981)1154.
\bibitem{Shiino} E. Shiino, Y. Saito, M. Ichimura and H. Toki,
Phys. Rev. {\bf C34}(1986)1004. 
\bibitem{Pace} A. De Pace, Phys. Rev. Lett. {\bf 75}(1995)29.
\end{thebibliography}
\end{document}